\documentclass[twocolumn]{rbef}
\usepackage{lipsum}

\usepackage{bbm}
\usepackage{subfig}

\usepackage{xcolor}
\definecolor{verde}{rgb}{0,0.5,0}
\usepackage{listings}
\titulocabecalho{Construção de uma Maquete do Sistema Solar com Controle de Temperatura para Alunos com Deficiência Visual}
\autorcabecalho{M. S. Almeida, J. N. M. Castro, W. T. Cruz e R. Q. Almeida}

\numeracao{xxxx}
\volume{xx}
\numero{xx}
\ano{2019}
\doi{http://dx.doi.org/xxxx}
\tipodeartigo{Produtos e Materiais Didáticos para o Ensino de Física}
	\author[1]{Maurício S. Almeida}
    \affil[1]{Departamento de Física, Instituto Federal do Ceará - campus Juazeiro do Norte 
    
      Av. Plácido Aderaldo Castelo, 1646 - Planalto, Juazeiro do Norte, CE, Brasil
      \thanks{\href{emailto:mauricio.almeida@ifce.edu.br}{mauricio.almeida@ifce.edu.br}}}
    \author[1]{João N. M. Castro}
    \author[1]{Wilami T. Cruz}
    \author[1]{Rodrigo Q. Almeida}

\titulo{Construção de uma Maquete do Sistema Solar com Controle de Temperatura para Alunos com Deficiência Visual}

\subtitulo{Construction of a Model of the Solar System with Temperature Control for Visually Impaired Students}

\begin{document}

\begin{primeirapagina}

%\begin{center}
%\vspace{-12pt}
%\small{Recebido em xxx. Aceito em xxx}
%\end{center}

	\begin{abstract}
Neste trabalho apresentamos um estudo realizado para o desenvolvimento de um conjunto didático baseado no uso do tato e receptores térmicos para o ensino de Astronomia.  A proposta surgiu diante das dificuldades encontradas na busca de experimentos capazes de apresentar da melhor forma conceitos envolvidos no ensino de Astronomia para pessoas com deficiência visual.  O equipamento consiste no uso de uma maquete com hemisférios dotados de pastilhas de Peltier para simulação de escala de temperatura de planetas do Sistema Solar.  A pesquisa baseou-se no uso de materiais acessíveis e de baixo custo. 
Todos os materiais utilizados e a montagem do equipamento  são descritas neste trabalho, desde o projeto até a aplicação do produto final para estudantes. Este experimento fez parte de um conjunto de equipamentos pedagógicos apresentados em uma mostra científica para deficientes visuais durante a 15\textsuperscript{a} Semana Nacional de Ciência e Tecnologia na cidade de Juazeiro do Norte no Estado do Ceará. 
    \palavraschave{Astronomia, Física, Ensino, deficientes visuais}
 
	\end{abstract}
	
	\begin{otherlanguage}{english}

	\begin{abstract}
In this work, we present a study carried out to develop a didactic kit based on the use of touch and thermal receivers for the teaching of Astronomy. The proposal arose in the face of the difficulties encountered in the search for experiments capable of presenting the best concepts involved in the teaching of Astronomy for visually impaired people. The equipment consists of the use of a model with hemispheres with Peltier pellets for simulation of temperature scale of planets of the Solar System. The research was based on the use of accessible and low-cost materials.
All the materials used and the assembly of the equipment are described in this work, since the design until the application of the final product for students. This experiment was part of a set of pedagogical equipment presented at a scientific exhibition for the visually impaired during the 15 \textsuperscript{a} National Science and Technology Week in the city of Juazeiro do Norte in the State of Ceará.
	
	\end{abstract}
	\end{otherlanguage}

	\end{primeirapagina}
\saythanks

\section{Introdução} \label{Introducao}

A intenção em pesquisar novas abordagens para o ensino de Astronomia surgiu a partir do exercício da docência para alunos deficientes visuais no campus Juazeiro do Norte do Instituto Federal do Ceará. A Instituição dispõe de um laboratório de Física bem equipado e um Núcleo de Astronomia responsável por ações de divulgação e popularização da ciência para a comunidade local. Também dispomos de um Núcleo de Atendimento às Pessoas com Necessidades Específicas (NAPNE)  que atua na inclusão de alunos e servidores com necessidades específicas no cotidiano escolar da instituição.  Apesar da boa estrutura física disponível aos docentes e da presença de um núcleo de apoio, a abordagem de alguns conceitos  no ensino de Física e Astronomia para alunos deficientes visuais sempre requerem  a adoção de técnicas inovadoras em sala de aula. O uso do braile constitui um importante meio para transmissão de conceitos a partir da leitura, no entanto é limitado quando se deseja apresentar percepções normalmente apresentadas sob o ponto de vista visual, tais como a noção de espaço, tamanho e temperatura.  Diante das dificuldades encontradas para abordar temas onde a visão é o meio usual para transferência dos conceitos em estudo, pensamos em desenvolver mecanismos capazes de apresentar uma abordagem alternativa às aulas com quadro branco e multimídia.

Escolhemos a Astronomia como cenário para nosso estudo por ser a ciência na qual a abordagem dos conceitos aos estudantes é essencialmente visual, sendo no campo teórico ou na realização de aula práticas, que quase sempre são realizadas com auxílio de telescópios. O uso de modelos em escala constitui um importante método apresentar noções de tamanho de planetas e distâncias no sistema solar e no universo. A aplicação de texturas  podem ser excelentes meios para apresentar a distribuição de estrelas em galáxias e aglomerados \cite{Weferling}. Para uma escala astronômica menor, os autores do trabalho \cite{Bernhard} propõem  o uso de modelos em duas e três dimensões para apresentar representações de asteroides próximos da Terra descobertos recentemente.

Neste cenário, a construção de kits didáticos bem como a possibilidade de inovação nas abordagens pedagógicas envolvendo experimentos, torna-se inviável e desestimulante devido ao custo, complexidade e baixa disponibilidade dos materiais utilizados para montagem.  Se pensarmos em escolas localizadas em cidades distantes dos grandes centros e capitais, este fato torna-se ainda mais agravante. Partindo desta realidade, a proposta de construção da ferramenta didática apresentada neste trabalho foi planejada a partir da aplicação de materiais e equipamentos de baixo custo e boa disponibilidade no comércio, como também feito por ouros autores \cite{Kitchin,Dominici,Lima,Beattie}.  Todos os materiais utilizados para montagem da parte estética do equipamento, tais como as esferas ilustrando planetas, a base para montagem e as texturas, foram adquiridas em papelarias e lojas de artesanato. Os insumos eletrônicos utilizados também foram comprados em lojas na cidade de Juazeiro do Norte, município do interior do Ceará. 

O presente estudo teve início no ano de 2018 quando o Ministério da Ciência, Tecnologia, Inovações e Comunicações (MCTIC) e o Conselho Nacional de Desenvolvimento Científico e Tecnológico (CNPq) lançaram a CHAMADA CNPq/MCTIC-SEPED Nº 14/2018 para seleção de propostas projetos de eventos de divulgação e popularização da ciência a serem realizados durante a Semana Nacional de Ciência e Tecnologia (SNCT 2018). O tema escolhido para a SNCT 2018 foi “Ciência para a Redução das Desigualdades”, de acordo com os Objetivos do Desenvolvimento Sustentável (ODS) estipulados pelas Nações Unidas.  Diante da oportunidade de captação de recursos do referido edital foi então concebida a proposta de realização de uma mostra científica com foco nas pessoas com deficiência visual com um conjunto de experimentos em que os participantes pudessem interagir tolhidos da visão por meio de vendas.  
Com a aprovação do projeto passamos à fase de elaboração dos protótipos dispondo de um curto intervalo de tempo até a data prevista para realização do evento.  Além disso, os recursos captados contemplaram apenas despesas de custeio nos limitando a aquisição de insumos e peças. 

Este trabalho é organizado na seguinte sequência: no capítulo 2 apresentamos uma breve revisão a respeito do efeito Peltier; no capítulo seguinte detalhamos todo esquema de montagem da eletrônica e da parte estética do experimento; no capítulo 4 apresentamos nossas considerações finais e perspectivas.

\section{Efeito Peltier} \label{Peltier}
Observado em 1834 pelo físico francês Jean Charles Athanase Peltier, o chamado efeito Peltier consiste na produção de um gradiente de temperatura em uma junção metálica (conhecido como termopar) de diferentes materiais, quando esses são submetidas a uma diferença de potencial. Esse fenômeno também é conhecido como o inverso do efeito de Seebeck, descoberto 13 anos antes, que descreve a produção de energia elétrica através de calor. Na verdade, os efeitos de Seebeck e Peltier são diferentes manifestações do mesmo processo físico \cite{Goldsmid:book}.

Uma das aplicações industriais são as chamadas pastilhas Peltier, que consistem em duas placas paralelas e isolantes, geralmente cerâmicas, onde no seu interior existem junções de semicondutores do tipo "n" e "p" conectados em uma matriz em que os pares possuem conexão elétrica em série e conexão térmica em paralelo. Sendo assim, quando uma corrente flui pelo sistema, ocasiona uma diferença de temperatura, fazendo com que haja a absorção de calor de um lado das placas e dissipação pelo outro lado. Um esquema simplificado é mostrado na Fig.

\begin{figure}[!htb]
\centering
\includegraphics[scale=0.32]{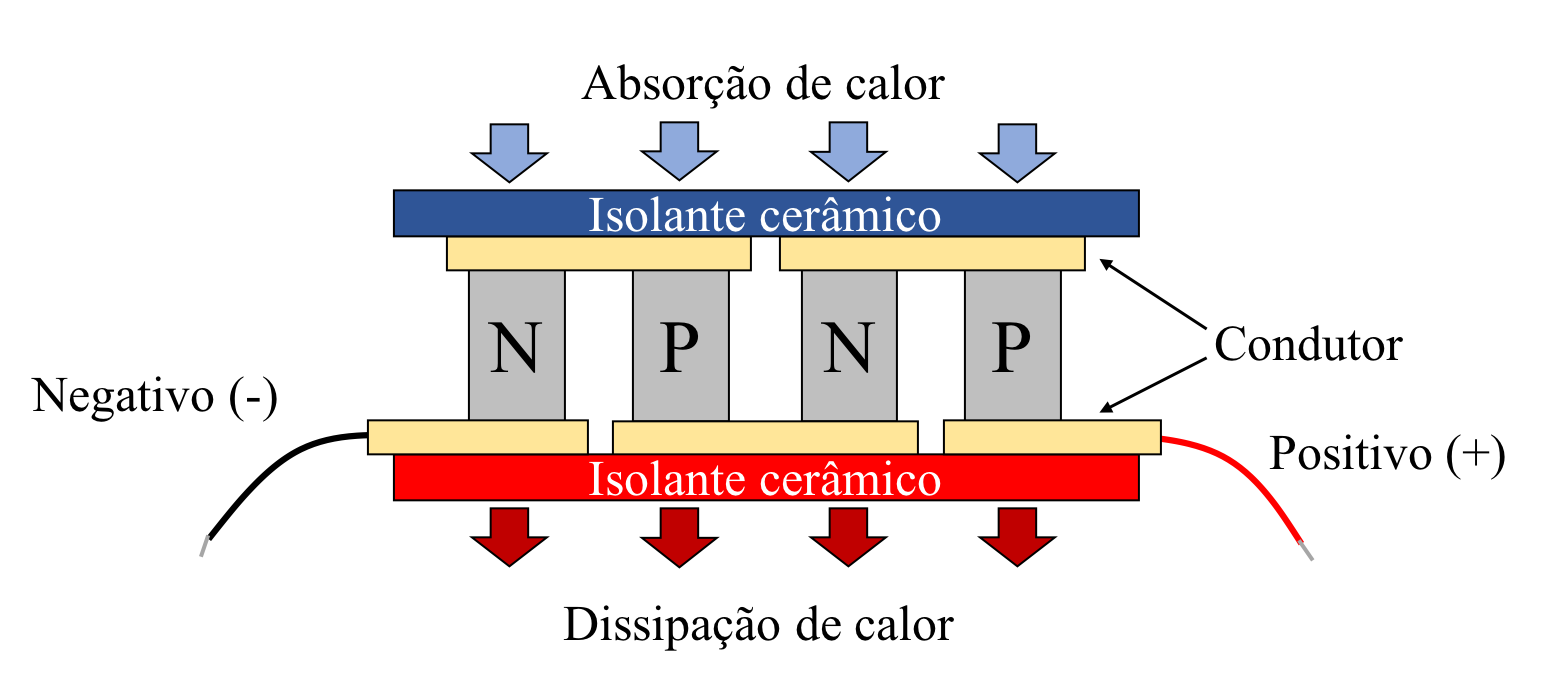}
\caption{Esquema de funcionamento de uma pastilha Peltier.}
\label{peltier}
\end{figure}

O efeito Joule talvez seja a mais conhecida interação entre um fenômeno elétrico e um fenômeno térmico associado. Ele é dado por,

\begin{equation}
Q=\int_{t_1}^{t_2}RI^{2}dt \label{EfeitoJoule}
\end{equation}%
onde $Q$ é a energia calorífica produzida devido a passagem da corrente, $R$ a resistência elétrica do condutor, $I$ a intensidade da corrente elétrica que poercorre o condutor no intervalo de tempo de $t_1$ a $t_2$. Quando a corrente é constante, a equação  (\ref{EfeitoJoule}) se resume a

\begin{equation}
Q = RI^{2}t \label{EfeitoJouleConstante}
\end{equation}%

Para o efeito Peltier, a energia dissipada ou absorvida $Q_p$ também é proporcional à corrente elétrica \cite{Patterson}:

\begin{equation}
Q_p = I\alpha\Delta T \label{EfeitoPeltier}
\end{equation}%
ou simplesmente,
\begin{equation}
Q_p = \pi\Delta T \label{EfeitoPeltier2}
\end{equation}%
sendo $\pi=I\alpha$ o coeficiente de Peltier, $\Delta T$ a diferença de temperatura entre o lado quente e o lado frio e $\alpha$ o coeficiente de Seebeck.
Dessa forma, quando se conecta um elemento do tipo Peltier a uma fonte de alimentação CC, a potência absorvida $P_a$ é decorrente do efeito Joule e do efeito Peltier, como representado a seguir

\begin{equation}
P_a = RI^{2}t + I\alpha\Delta T \label{PotenciaFonte}
\end{equation}%

\section{Montagem Experimental} \label{Montagem}

Optou-se para a montagem do projeto a utilização de materiais de fácil acesso e manuseio, para que pudesse ser replicado por um maior número de pessoas. Os materiais adotados foram:
\begin{itemize}
  \item 2 folhas de isopor de dimensões $70\times50\times3$\,cm;
  \item 4 bolas ocas de isopor com diâmetros de 5.0, 7.5, 20 e 25\,cm; 
  \item 2 bolas ocas de isopor com diâmetro de 15\,cm; 
  \item 2 bolas ocas de isopor com diâmetro de 10\,cm;
  \item 7 pastilhas termoelétricas Peltier TEC1-12706, $40\times40$\,mm;
  \item 5 coolers 12\,V DC + dissipadores de calor;
  \item 1 fonte 12\,V, 6\,A;
  \item 1 Folha de EVA;
  \item Pasta térmica;
  \item Fios e conectores;
  \item Tinta Guache (branco, preto, verde, azul, amarelo e vermelho);
  \item Cola de isopor, cola quente, algodão e folha de acetato transparente.
\end{itemize}

A base da maquete foi feita juntando as folhas de isopor, totalizando uma área útil de $140\times 50$\,cm$^{2}$. As bolas foram utilizadas na representação dos planetas; para tanto foram divididas ao meio e fixadas com cola de isopor sobre a base de forma alinhada e igualmente espaçadas, para facilitar a percepção através do tato. 

Por não haver um grande espectro de diâmetros de bolinhas de isopor nas papelarias e para tentar retratar que existem planetas com tamanhos aproximados, optou-se por usar os hemisférios das bolas de 10\,cm de diâmetro para representarem a Terra e Vênus, bem como também foi adotado os hemisférios das bolas de 15\,cm de diâmetro para os planetas Urano e Netuno. Os hemisférios de 5.0, 7.5, 20 e 25\,cm foram escolhidos para a ilustração de Mercúrio, Marte, Saturno e Júpiter, respectivamente.

Para servir como referencial, um arco de circunferência foi feito com a folha de EVA e colada na maquete para representar a posição do Sol. Optou-se por esse material diferente para não confundir com os planetas. 

\begin{figure}[!htb]
\centering
\includegraphics[scale=0.8]{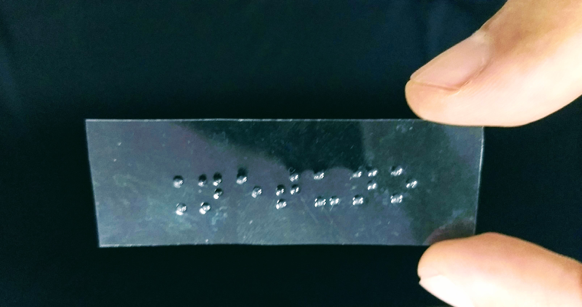}
\caption{Legenda do planeta Netuno feita em papel acetato transparente. Esse material é indicado por possuir boas propriedades mecânicas, fazendo com que a impressão não se desfaça com facilidade.}
\label{netuno}
\end{figure}

Após a montagem estrutural do experimento, sua pintura e secagem, colou-se legendas em Braille feitas em folha de acetato transparente (Fig. \ref{netuno}), indicando os respectivos planetas. A escolha por essa folha é a sua alta durabilidade, quando comparada com outros tipos de materiais como o papel ou cartolina, por exemplo.

\begin{figure}[!htb]
\centering
\includegraphics[scale=1.0]{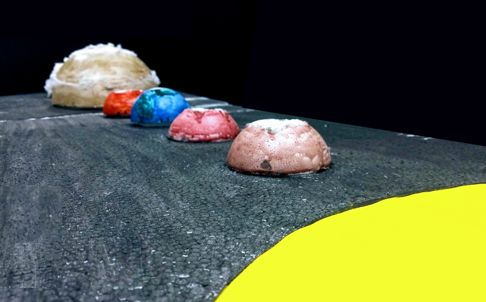}
\caption{Parte da maquete do Sistema Solar. Os hemisférios devem estar alinhados, para que os alunos possam facilmente localiza-los durante o manuseio.}
\label{sistemaSolar}
\end{figure}

Para diferenciar os planetas gasosos (Júpiter, Saturno, Urano e Netuno) dos demais, colou-se sobre eles pequenos pedaços de algodão, para proporcionar uma textura diferente. A Fig. \ref{sistemaSolar}  apresenta a montagem da maquete do Sistema Solar.

\begin{figure}[!htb]
\centering
\includegraphics[scale=0.9]{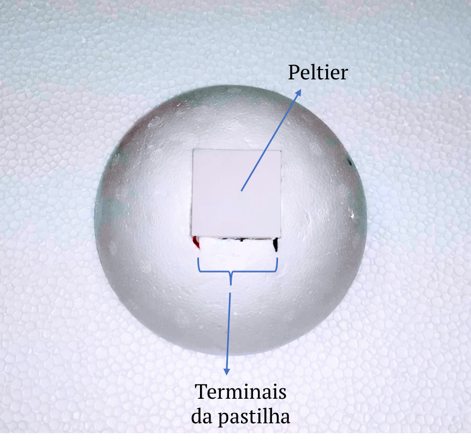}
\caption{Procedimento para fixação da pastilha de Peltier. Os fios devem ficar na parte interna do hemisfério de isopor. Para melhorar a aderência, foi aplicada uma fina camada de cola quente nas bordas da pastilha.}
\label{montagem}
\end{figure}

Uma abertura na parte superior dos hemisférios de isopor foi feita, com dimensão de $4\times4$\,cm, para a fixação das pastilhas termoelétricas (Fig. \ref{montagem}). Essas por sua vez, são as responsáveis pelo controle de temperatura do experimento. Os fios existentes foram escondidos na parte interna das bolas de isopor. Para melhorar a vedação, esse procedimento foi feito com cola quente.

\subsection{Controle de temperatura}
\begin{figure*}[!htb]
\centering
\includegraphics[scale=0.9]{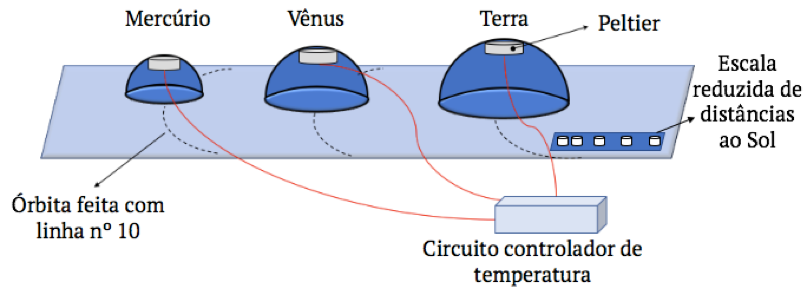}
\caption{Modelo de representação do Sistema Solar com controle de temperatura. A Peltier promoverá a sensação de que os planetas mais próximos ao Sol estejam mais aquecidos e que os mais afastados estejam mais gelados. Um cuidado especial deve se ter em relação a temperatura de Mercúrio, porque embora ele esteja mais próximo do Sol, possui temperatura média inferior à de Vênus.  Além disso, no canto inferior direito da maquete foi criada uma representação, em escala reduzida, das distâncias dos planetas ao Sol.}
\label{esquema}
\end{figure*}

A percepção da temperatura é feita ao tocar a pastilha que fica na parte superior de cada planeta, conforme apresentado no esquema da Fig. \ref{esquema}. Para efeito didático, buscou-se fazer uma escala linear do aumento de temperatura a medida que se aproxima do Sol. Por esse motivo, a Peltier do planeta Terra deve ser desligada para representar a temperatura ambiente. Um cuidado especial foi tido ao representar as temperaturas de Mercúrio e Vênus, uma vez que a temperatura deste é maior do que a daquele, embora Mercúrio seja o planeta mais próximo ao Sol. Esse ponto é importante, porque abre espaço para explicar as consequências do efeito estufa gerada pela pesada atmosfera de dióxido de carbono.

Para mostrar a distância média dos planetas ao Sol, foi criada, ainda na maquete da Fig. \ref{esquema}, pequenas marcações em formato esférico, com cola de relevo 3d, para representarem a posição dos planetas do Sistema Solar. A Tab. \ref{tabela1} mostra a distância de cada uma delas ao ponto de referência (Sol). Adotou-se que cada unidade astronômica correspondia a 1\,cm. Como a espessura das marcações era da ordem de mm, foi necessário fazer alguns arredondamentos.

\begin{table}[!htb]
  \centering
  \begin{tabular}{cc}
    \hline
    Planetas & Distância em relação ao Sol (cm) \\
    \hline
    Mercúrio & 0,4\\
    Vênus & 0,7\\
    Terra & 1,0\\
    Marte & 1,5\\
    Júpiter & 5,0\\
    Saturno & 9,5\\
    Urano & 19\\
    Netuno & 30\\
    \hline
  \end{tabular}
  \caption{Distâncias adotadas para a representação em escala reduzida das distâncias médias dos planetas ao Sol.}
  \label{tabela1}
\end{table}

\subsubsection{Controle de temperatura por arduino
}
O controle de temperatura da pastilha Peltier pode ser facilmente implementado utilizando Arduino (que é uma plataforma de prototipagem eletrônica) ou simplesmente com um microcontrolador, através da técnica de PWM (\textit{Pulse Width Modulation}) onde uma onda quadrada é criada, alternando assim o sinal entre ligado e desligado. A duração do "\textit{on-time}" é chamada de largura do pulso.

\begin{figure}[!htb]
\centering
\includegraphics[scale=0.7]{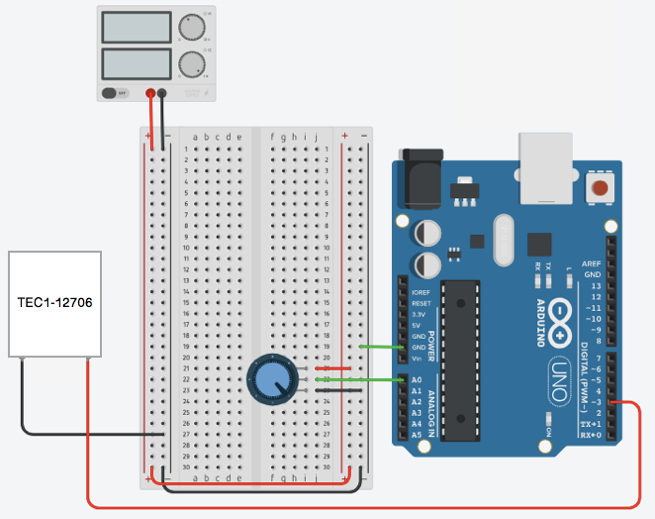}
\caption{Circuito para controle de temperatura da Peltier utilizando PWM. Foi utilizado um Arduino Uno R3, um potenciômetro de 1\,k$\Omega$ e 5\,W, fonte de alimentação 12\,V 6\,A e uma TEC1-12706. Além disso, também é necessário um sistema de dissipação de calor (não mostrado na figura) para a parte quente da pastilha. Para aplicações em que se deseje um maior aquecimento/resfriamento, deve-se substituir o potenciômetro por um de resistência menor.}
\label{arduino}
\end{figure}

Um circuito simplificado para apenas uma pastilha é mostrado na Fig. \ref{arduino}. A largura do pulso é modulada pelo potenciômetro de 1\,k$\Omega$. Para essa situação foi utilizada uma fonte de 12\,V e 6\,A. Dependendo da corrente desejada, também é aconselhável fazer uso de um Mosfet canal N e um resistor de 10\,k$\Omega$.

As portas digitais 3, 5, 6, 9, 10 e 11 do Arduino são destinadas ao PWM. Em nosso exemplo, foi adotada a porta 3 para este fim. O código utilizado para o controle de temperatura das pastilha Peltier é mostrado a seguir:

\begin{lstlisting}
int peltier = 3;
void setup(){
  Serial.begin(9600);
  pinMode(A0, INPUT);
  pinMode(peltier, OUTPUT);
}
void loop(){
  analogWrite(peltier, map(analogRead(A0), 0, 1023, 0, 255));
  delay(10);
Serial.print(" 0 ");
Serial.print(" 255 ");
Serial.println(map(analogRead(A0), 0, 1023, 0, 255)); 
 } \end{lstlisting}

Para melhorar o desempenho das pastilhas que serão responsáveis pela temperatura dos planetas frios, é necessário acoplar na parte quente de cada uma delas o sistema de dissipação de calor formado pelo cooler mais dissipador, caso contrário a Peltier entrará em equilíbrio térmico impedindo a percepção do frio. Esse procedimento, entretanto, não é necessário ser feito para as pastilhas que ficarão em Mercúrio e Vênus, uma vez que para esses casos a parte quente ficará direcionada para cima, onde os alunos deverão tocar.

\subsubsection{Controle de temperatura com resistores de fio}
Como deseja-se produzir apenas um efeito didático, as pastilhas não precisam trabalhar em condições extremas e isso facilita a montagem do circuito, porque com pouca corrente já se obtém o efeito de aquecimento/resfriamento. Além disso, temperaturas elevadas poderiam causar desconforto ou até danos aos alunos.

\begin{figure}[!htb]
\centering
\includegraphics[scale=0.85]{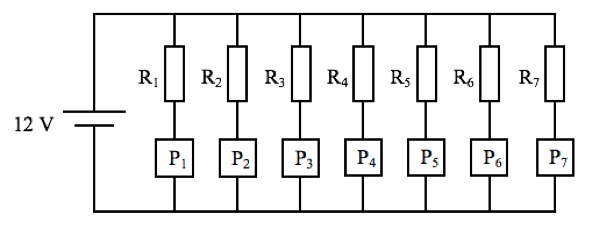}
\caption{ Circuito controlador de temperatura. As resistências R$_1$, R$_2$, ... R$_7$ são de 40, 60, 120, 60, 40, 30 e 20 $\Omega$, respectivamente. As pastilhas P$_1$, P$_2$, ... P$_7$, são referentes aos planetas Mercúrio, Vênus, Marte, Júpiter, Saturno, Netuno e Urano, respectivamente. Adotou-se uma fonte de 12\,V e 6\,A. A figura não mostra a conexão dos dissipadores de calor.}
\label{resistores}
\end{figure}

A Fig. \ref{resistores} mostra o esquema do circuito para controle de temperatura. Recomenda-se a utilização de resistores de fio com potência mínima de 5 Watts, que são capazes de realizar o controle de corrente elétrica sem queimar. Similarmente ao caso onde é empregado o Arduino, também é necessário a utilização do sistema de dissipação de calor, com coolers, nas pastilhas que representarão os planetas frios.

\section{Considerações Finais} \label{Consideracoes}
\begin{figure}[!htb]
\centering
\includegraphics[scale=0.65]{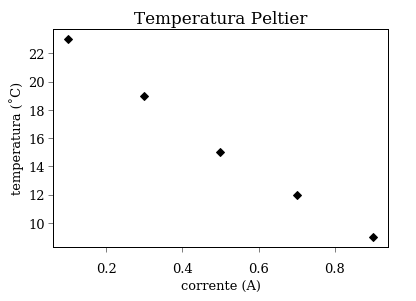}
\caption{ Temperatura da pastilha termoelétrica Peltier TEC1-12706 em função da corrente. Foi adotado uma fonte de 12 Volts nesse ensaio.}
\label{temperatura}
\end{figure}

A Fig. \ref{temperatura} mostra a temperatura obtida pelas pastilhas em função da corrente do aplicada. Para esse ensaio foi utilizado uma fonte regulável de 12\,V e 6\,A. A temperatura foi aferida com um termômetro laser digital.

Essa maquete, juntamente com outros modelos experimentais, foram apresentados em uma mostra de astronomia durante a Semana Nacional de Ciência e Tecnologia (SNCT) 2018, no IFCE campus Juazeiro do Norte. Para isso foi reservado uma sala, que foi completamente vedada, onde os visitantes que não tinham deficiência visual entravam vendados para conhecer as maquetes e terem a mesma experiência dos deficientes visuais. 

É necessário empenhar esforços para
desenvolver mecanismos de aprendizagem que contribuam para o ensinamento da
física e da astronomia para alunos com deficiência visual, de tal maneira que eles também
possam, através do tato, contemplar a beleza do Universo. Bavcar, filósofo esloveno, já dizia: “...o teu horizonte é até onde você pode ver. Se você vê com as mãos, logo o teu horizonte é até onde você pode tocar”.

Essa maquete pode ser replicada com facilidade e demonstrou ser um bom recurso didático, capaz de tornar paupável conceitos relevantes sobre o nosso Sistema Solar e temas transversais ao assunto.

%\section*{Agradecimentos}

\end{document}